# Atomic Visualization of Bulk and Surface Superconductivity in Weyl Semimetal $\gamma$-PtBi$_2$

Hao Zhang[1,2†], Hui Chen[1,2,3†*], Zichen Huang[1,2], Zi-Ang Wang[1,2], Guangyuan Han[1,2], Ruisong Ma[1,2], Xiangde Zhu[4], Wei Ning[4], Chengmin Shen[1,2], Qing Huan[1,2], and Hong-Jun Gao[1,2,3*]

[1] *Beijing National Center for Condensed Matter Physics and Institute of Physics, Chinese Academy of Sciences, Beijing 100190, PR China*

[2] *School of Physical Sciences, University of Chinese Academy of Sciences, Beijing 100190, PR China*

[3] *Hefei National Laboratory, 230088 Hefei, Anhui, PR China*

[4] *Anhui Province Key Laboratory of Condensed Matter Physics at Extreme Conditions, High Magnetic Field Laboratory, Chinese Academy of Sciences, Hefei, 230031, China*

[†] These authors contributed equally to this work

[*] Corresponding authors: hjgao@iphy.ac.cn, hchenn04@iphy.ac.cn

**Abstract**

A bulk superconductor hosting intrinsic surface superconductivity provides a unique platform to study Majorana bound states. The superconductor, trigonal $\gamma$-PtBi$_2$, is a promising candidate, as surface superconducting gaps and topological surface states have been observed. However, the simultaneous presence of bulk and surface superconductivity has not been resolved. Here, we directly visualize coexisting bulk and surface superconducting gaps in trigonal PtBi$_2$ by using ultra-low-temperature scanning tunneling microscopy/spectroscopy. The bulk gap is $\Delta \sim 0.053$ meV with a critical temperature ($T_c$) $\sim 0.5$ K and a critical field below 0.01 T, accompanied by a vortex lattice and bound states, yielding a coherence length of $\sim$200 nm. Remarkably, certain surface regions show a much larger gap of $\Delta \sim 0.42$ meV, persisting up to $T_c \sim 3$ K and surviving magnetic fields up to 2 T, yet lacking a static vortex lattice. This coexistence of robust surface and bulk superconductivity establishes $\gamma$-PtBi$_2$ as a unique platform for investigating the interplay between bulk and surface Cooper pairings in superconducting topological materials.

## Introduction

Superconducting topological materials have emerged as a forefront topic in condensed matter physics[1,2]. Despite the growing interest, only a limited number of candidate systems[3,4] have been identified because experimentally verifying topological superconductivity remains highly challenging. To overcome this difficulty, numerous experimental efforts have focused on inducing superconductivity on the surfaces of topological materials, including the fabrication of heterostructures[5–7] and the use of tip- or proximity-induced superconductivity[8–10]. A key breakthrough has been the discovery of bulk single-crystal superconductors that host topological surface bands, which can be regarded as intrinsic heterostructures[11–16]. Moreover, the topological surface states of non-superconducting three-dimensional (3D) semimetals have been shown to exhibit intrinsic surface superconductivity below their critical temperature, where the spin–momentum-locked Fermi surfaces may also give rise to topological superconductivity[17,18]. In rare cases, a bulk superconductor that intrinsically hosts superconducting surface states[19,20], which allows bulk and surface Cooper pairings to coexist, offers a unique platform for exploring the interplay between superconductivity and topology[21].

$PtBi_2$ is one of such candidates showing traces of superconductivity as well as the presence of Dirac-like states and triply degenerate Weyl points near the Fermi level. Semimetallic $PtBi_2$ is known to exist in polymorphic forms with cubic and trigonal structures. The cubic-$PtBi_2$ ($\beta$-$PtBi_2$) shows extremely large unsaturated magnetoresistance[22,23], while the trigonal $PtBi_2$ ($\gamma$-$PtBi_2$) was predicted to possess type-I Weyl semimetal band structures[24]. Due to its strong spin−orbit coupling and inversion symmetry breaking, a variety of interesting electronic properties have been reported, including strong Rashba-like spin splitting[25], one-dimensional gap-protected edge states at single-layer steps[26], pressure-induced superconductivity[27] and tip-enhanced superconductivity[28,29]. Recent experiments have hinted at independent intrinsic superconducting behaviors in the bulk and on the surface of $\gamma$-$PtBi_2$. Transport measurements on bulk single crystals and thin films showed superconductivity with an ultra-low critical temperature of <1 K[30,31]. The Berezinskii-Kosterlitz-Thouless behavior is observed in surprisingly thick films, suggesting that the superconductivity may be of 2D origin. Later, angle-resolved photoemission spectroscopy (ARPES) studies reported surface superconductivity with a $T_c$ of around 10 K and a superconducting gap of ~1 meV where the bulk remained in the normal state[32–34]. Although a surface superconducting gap[35–37] and topological surface states[38,39] have been recently reported by scanning tunneling microscopy (STM), the gap size and spatial inhomogeneity of the observed surface superconducting gap are unusually large. In addition, the real-space study of coexistence of bulk and surface superconductivity has not been reported.

In this Letter, we report the atomic visualization of the bulk and surface superconducting gaps, using an ultra-low temperature (5 mK) scanning tunneling microscope/spectroscopy. The bulk superconducting gap has a size of $\Delta$ ~ 0.053 meV, with a critical temperature of approximately 0.5 K and a critical field below 0.01 T. Zero-energy conductance maps under a magnetic field reveal bulk superconducting vortices and vortex bound states, with an estimated coherence length of ~200 nm. Remarkably, in some surface regions, we observe a larger gap with a size of $\Delta \sim 0.42$ meV below a higher critical temperature of ~3 K and a stronger critical field of ~2 T. In contrast to the bulk superconducting gap, no static vortex lattice can be resolved for this larger surface gap. The new superconducting gap is robust against variations in surface termination and intrinsic defects. Moreover, the Fermi arcs exhibit no discernible difference between surface regions with and without the surface superconducting gap. Our findings provide a new insight into the surface and bulk superconductivity in the type-I Weyl semimetal.

## Results and Discussion

Topological band structure arises in $\gamma$-PtBi$_2$ due to the breaking of inversion symmetry in its layered lattice structure, as illustrated in [Fig. 1(a)], where each layer is composed of one layer of platinum atoms sandwiched between two layers of bismuth atoms with different arrangements. Consequently, there are two types of bismuth-terminated cleavage surfaces [Fig. 1(a)], one possessing a decorated honeycomb structure (*H*-type) while the other is Kagome-like (*K*-type). Both types of cleavage planes are clearly resolved by STM, showing excellent agreement with the ideal crystal structure [Fig. 1(b)].

We first study the density of states on the surface of $\gamma$-PtBi$_2$ by collecting the differential conductance (d$I$/d$V$) spectra at an ultra-low base temperature of ~5 mK (details see Experimental Method section). A narrow energy gap symmetric about the Fermi level is observed in the spatially averaged d$I$/d$V$ spectrum across the entire surface of PtBi$_2$ [Fig. 1(c)]. Utilizing the Dynes model, which describes the narrow gap seamlessly [Fig. 1(c)], the gap size $\Delta$ is estimated to be approximately 0.053 meV. As the temperature increases, we find that the superconducting gap is fully suppressed at 0.5 K [Fig. 1(d)]. Furthermore, the temperature evolution of the superconducting gap size roughly agrees with the Bardeen-Cooper-Schrieffer (BCS) theory, yielding a critical temperature $T_c = 0.35$ K [Fig. 1(e)]. The gap size and critical temperature obtained from the STM measurements are consistent with the bulk superconductivity measured in transport experiments on $\gamma$-PtBi$_2$[30,31].

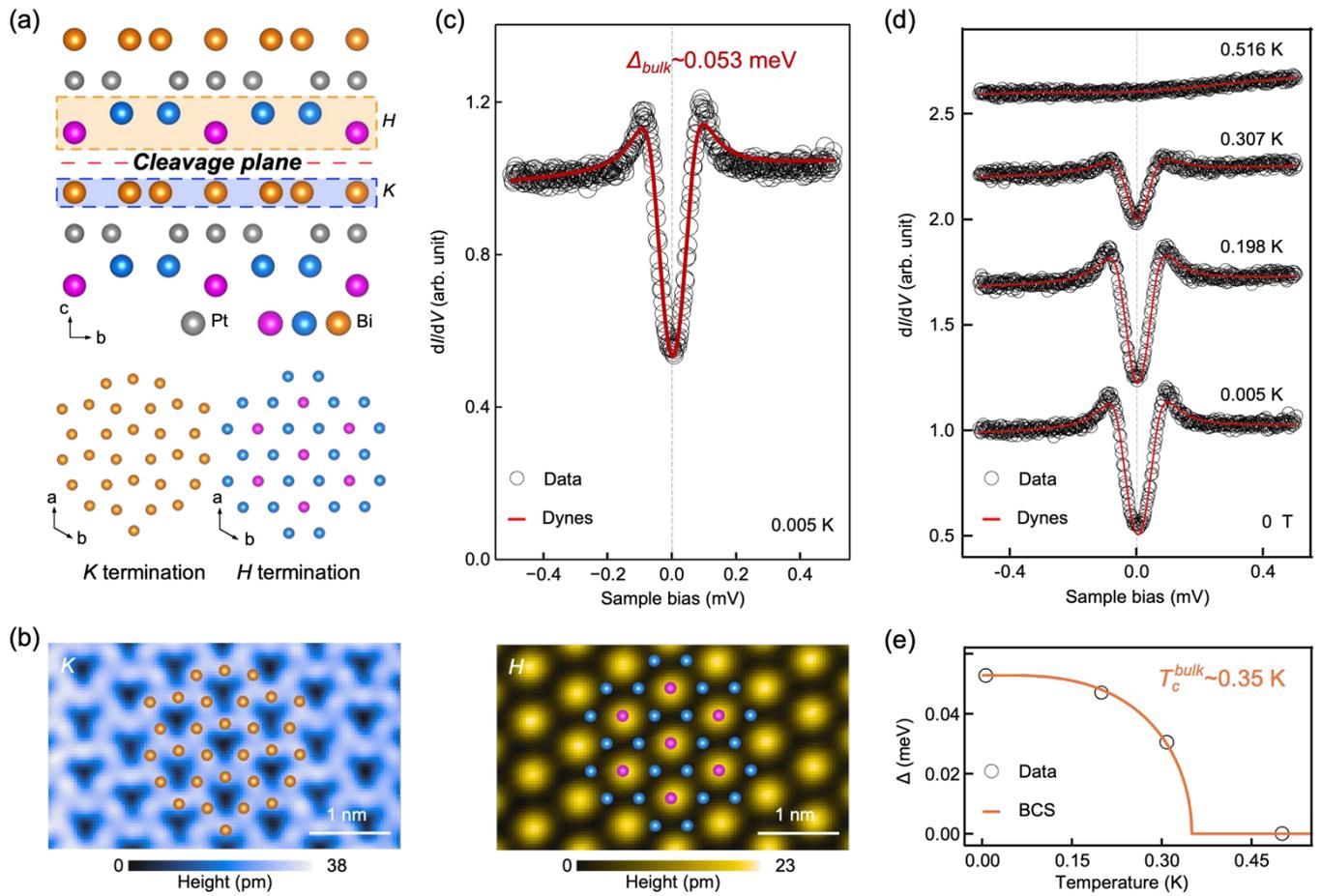

Fig. 1. Atomic model, cleaved surfaces and bulk superconducting gap of γ-PtBi$_2$. (a) Atomic model showing the non-centrosymmetric crystal structure of PtBi$_2$, which possesses two kinds of cleavage surfaces, annotated as H (decorated honeycomb) and K (Kagome), respectively, in this illustration. (b) Atomically resolved STM images of the $K$-type (left) and $H$-type (right) cleavage surfaces with the atomic arrangements of each type superimposed (setpoint: $V_s = -10$ mV, $I_t = 1$ nA). (c) Spatially averaged d$I$/d$V$ spectrum over the cleavage surface of PtBi$_2$, displaying a narrow superconducting gap. The red curve represents the fitting using the Dynes equation. (d) and (e) Evolution of the narrow superconducting gap as a function of temperature. Spectra at different temperatures are fitted utilizing the Dynes model (red curves in (d)) to obtain the temperature evolution of the gap size (e). The orange curve in (e) represents the behavior of the gap size at finite temperatures as predicted by the BCS theory. Setpoint in (c) and (d): $V_s = -2$ mV, $I_t = 1$ nA, $V_{mod} = 0.005$ mV.

We also reveal vortices and vortex bound states of the narrow superconducting gap by applying a magnetic field perpendicular to the surface of $\gamma$-PtBi$_2$. As the critical field is exceptionally low ($\mu_0 H_{C2}$ < 10 mT) [Fig. 2(a)], the narrow gap is fully suppressed by such weak out-of-plane magnetic fields that it is challenging to observe vortices directly. After reducing the magnetic field back to 0 T from relatively higher fields, the weak remnant field in the STM system enables the imaging of vortices. As shown in [Fig. 2(b)], a single isotropic magnetic vortex is observed in the zero-energy conductance map. The spatial evolution of d$I$/d$V$ spectra from the vortex center to the vortex halo displays the gradual closing of the superconducting gap [Fig. 2(c) and 2(d)]. Even at the center of the vortex, a weak gap is still observed in the d$I$/d$V$ spectrum [Fig. 2(d)]. These features are attributed to the spatial evolution of vortex bound states, such as Caroli-de Gennes-Matricon (CdGM) states, of the bulk superconductivity in a non-quantum-limit condition[40,41]. The small dip structure at the vortex core may result from the vortex bound states[42,43] or a "pseudogap"[44], as similar dip structures have been reported in the vortex cores of unconventional superconductors.

The ultra-small critical field of the narrow gap observed in STM measurements is consistent with the transport results[30,31]. The consistency of both the critical temperature and the critical field provides direct evidence that the narrow gap originates from the bulk superconductivity. Such agreement between STM and bulk transport measurements has been widely recognized as a criterion for identifying bulk superconductivity[45]. The spatial size of a vortex reflects the coherence length of bulk superconductivity. In [Fig. 2(e)], the zero-energy conductance retrieved from the d$I$/d$V$ linecut across the vortex center is fitted to an exponential decay, which gives an estimated coherence length for the bulk superconductivity of about $\xi \sim 200$ nm. Thus, we provide the first real-space observation of the bulk superconducting gap at an ultra-low temperature.

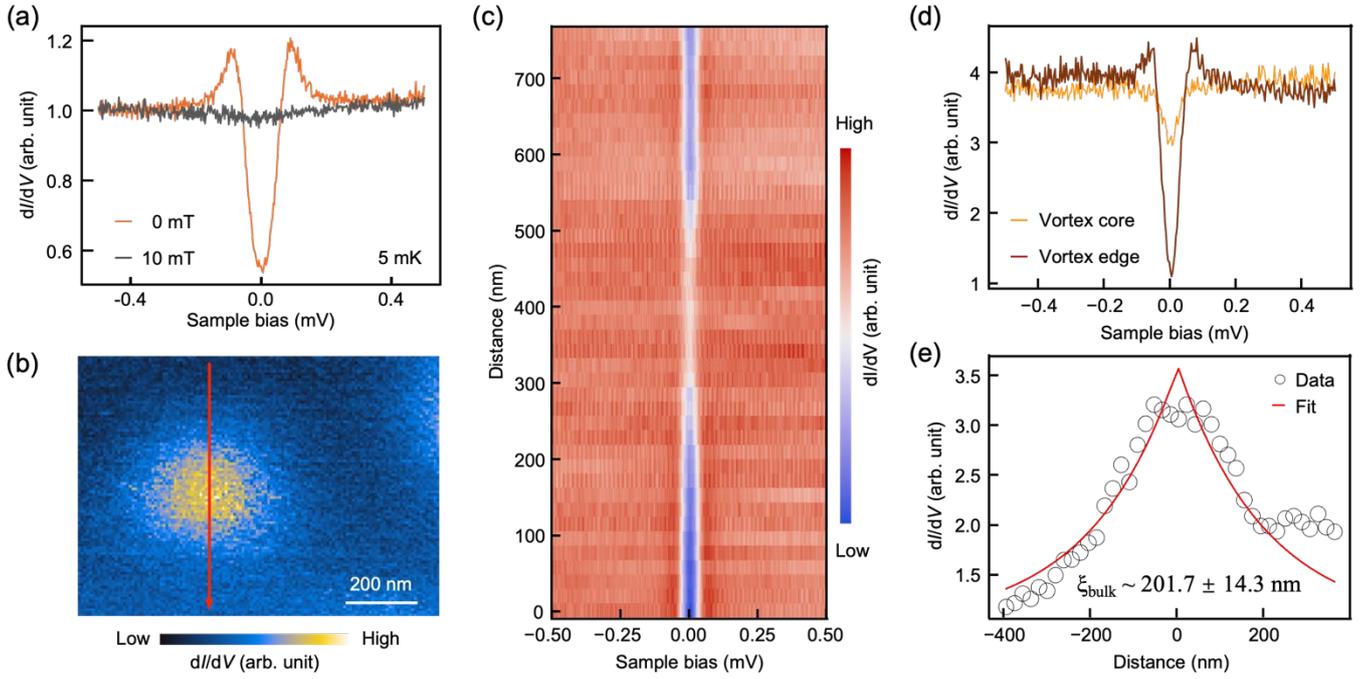

Fig. 2. Bulk superconducting vortices and vortex bound states in PtBi$_2$ under out-of-plane magnetic fields. (a) d$I$/d$V$ spectra under zero field and an out-of-plane magnetic field of 10 mT, respectively, showing the exceptionally low critical field of the bulk superconductivity. (b) Zero-energy conductance map showing a single vortex on the surface of PtBi$_2$. (c) Evolution of d$I$/d$V$ spectra across the vortex center along the red arrow marked in panel (b). (d) Comparison of d$I$/d$V$ spectra taken at the center of a vortex (orange) and far from the vortex (red), showing the suppression of the superconducting gap at the vortex core. (e) Intensity of normalized zero-energy conductance as a function of distance across the vortex core. Red curves represent the exponential fitting results, yielding a coherence length for the bulk superconductivity of about 200 nm. Setpoint in (a), (c) and (d): $V_s = -1$ mV, $I_t = 1$ nA, $V_{mod}$=0.005mV.

The possible topological surface superconductivity is the most intriguing nature of PtBi$_2$[32]. However, previous works reveal that surface superconductivity, with a much larger gap size than that observed in the ARPES work[32], is inhomogeneous in space and exhibits abnormal behaviors under increasing temperature and magnetic field[35,36]. Here, in addition to the first STM clue on the bulk superconductivity, possible surface superconductivity with a much larger gap size $\Delta_{surface} \sim 0.42$ meV in d$I$/d$V$ spectra [Fig. 3(a)] is observed in some surface regions. The evolution of the large energy gap with temperature and magnetic field provides estimates of the critical temperature $T_c \sim 3$ K [Fig. 3(b)] and the upper critical field $\mu_0 H_{C2} \sim 2$ T [Fig. 3(e)]. These values are considerably larger than those of the bulk superconductivity in the same sample discussed previously, excluding proximity-induced superconductivity from the bulk. In addition, the measured gap-to-$T_c$ ratio $2\Delta_{surface}/k_B T_c \sim 3.2$, smaller than that of the bulk superconductivity with the conventional weak-coupling strength of 3.5, further supports a distinct origin of the larger superconducting gap from the bulk superconductivity.

A possible origin of the large superconducting gap is convolution with a superconducting tip, which may occur if the STM tip accidentally picks up superconducting nanoflakes from the sample surface[46]. In such a scenario, a Josephson junction would form between the sample and the tip, producing a characteristic zero-bias peak in the d$I$/d$V$ spectra under sufficiently high tunneling conductance. However, as shown in Fig. 3(c), no signature of Josephson tunneling is observed even at tunneling currents up to 100 nA. In addition, the large superconducting gap persists across different tip conditions. These observations rule out a superconducting tip as the origin of the large gap $\Delta_{\text{surface}}$.

At zero magnetic field, the superconducting gap $\Delta_{\text{surface}}$ remains spatially uniform over regions extending several hundred nanometers. The d$I$/d$V$ spectra acquired using the same tip condition over various surface areas can be classified into two distinct types: the majority showing a bulk superconducting gap of approximately 0.05 meV, and a minority exhibiting a much larger gap of about 0.4 meV (Fig. 3(d)). This bimodal distribution indicates that the gap variation does not originate from a continuous spatial inhomogeneity but rather reflects the coexistence of two distinct superconducting phases.

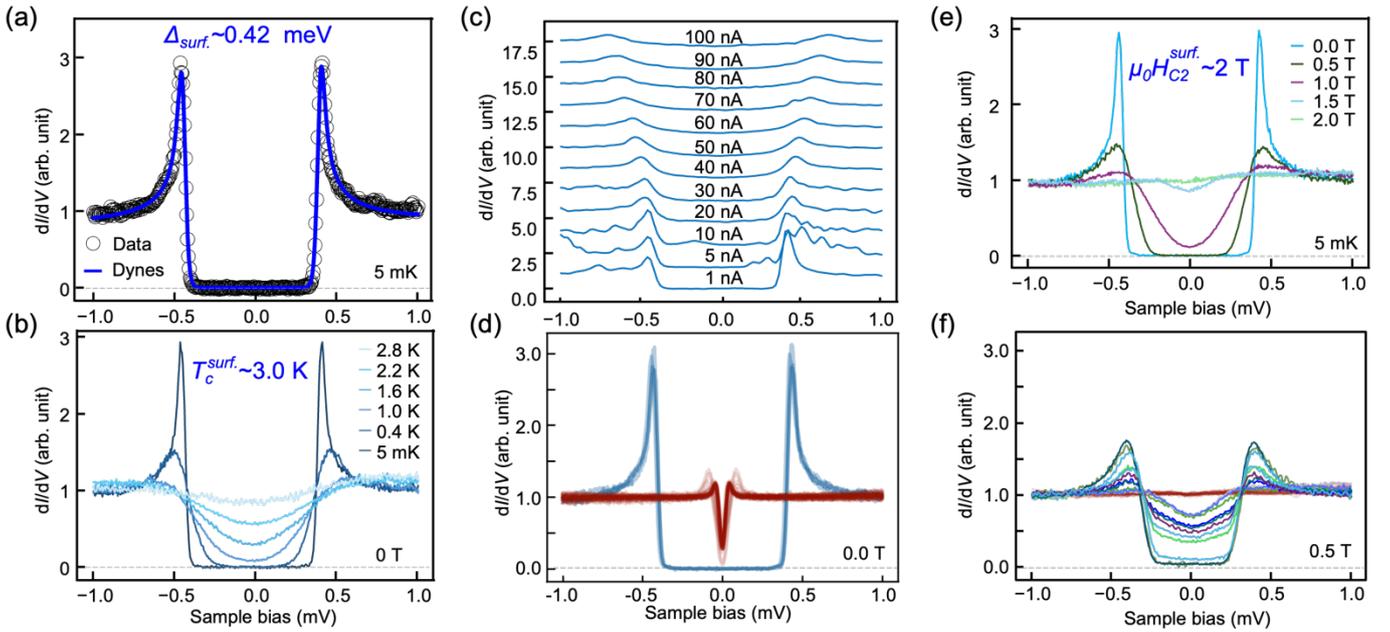

Fig. 3. Observations of surface superconducting gap and its evolution with temperature and magnetic fields. (a) A typical d$I$/d$V$ spectrum displaying a superconducting gap $\Delta$ = 0.42 meV which is much larger than the bulk superconducting gap shown in Fig. 1. The spectrum can be well described by the Dynes model (blue curve). (b) Evolution of the d$I$/d$V$ spectra with temperature, which gives an estimation of the superconducting critical temperature around 3 K. (c) Evolution of d$I$/d$V$ spectra with increasing tunneling conductance, showing the absence of the Josephson peak at zero energy and ruling out the possibility of superconducting STM tip ($V_s$ = −1 mV, $V_{\text{mod}}$=0.005 mV). (d) Stack plots of a series of d$I$/d$V$ spectra over various surface regions at zero field, showing large-gap surface regions (blue color) and small-gap surface regions (red color). (e) Evolution of d$I$/d$V$ spectra with out-of-plane magnetic fields, showing that the surface superconducting gap is fully suppressed under a critical field of 2 T. (f) Stack plots of a series of d$I$/d$V$ spectra over various surface regions, the same as in (d) but under a 0.5 T magnetic field, showing similar behavior of vortex bound states. Setpoint in (a-b), (d-f): $V_s$ = −1 mV, $I_t$ = 1 nA, $V_{\text{mod}}$=0.005 mV.

The magnetic vortex is another significant subject in topological superconductors for its potential in hosting novel topological states. Though the wide superconducting gap is not fully destroyed until a 2 T out-of-plane magnetic field [Fig. 3(e)], magnetic vortices are absent in the zero-energy d$I$/d$V$ maps under the magnetic fields < 2 T. Nonetheless, in contrast to the similar gap size across various surface regions at zero field [Fig. 3(d)], the d$I$/d$V$ spectra under a magnetic field of $\mu_0 H$ = 0.5 T show considerable variation in gap depth in different regions, akin to the variation of the superconducting gap inside a vortex [Fig. 3(f)]. It is worth noting that the d$I$/d$V$ spectra in each surface region are identical despite their alternation between regions. These experimental findings suggest that the surface superconducting state may host dynamically induced vortex states activated by the STM tip[47].

Although both bulk and surface superconducting gaps are clearly visualized at atomic scale, the origin of the spatially confined surface superconductivity remains elusive. First, the type of surface termination cannot account for the area dependence: a surface superconducting gap is observed on both cleavage terminations and remains uniform across their boundaries [Fig. 4(a) and 4(b)]. Surface defects are also unlikely to play a role, as the d$I$/d$V$ spectra acquired on and away from defects are nearly identical [Fig. 4(c) and 4(d)], demonstrating the robustness of surface superconductivity against local disorder. On the other hand, surface Fermi arcs are considered the key ingredient for the formation of surface superconductivity[32]. The absence of Fermi arcs would, in principle, lead to its disappearance. However, as demonstrated in [Fig. 4(f) and 4(h)], quasiparticle interference measurements near the Fermi level reveal scattering signatures between Fermi arcs[38] in both regions with and without surface superconductivity. These observations suggest that, in addition to Fermi arcs, other mechanisms must contribute to the emergence of surface superconductivity, warranting further investigation.

Recent STM[35,36] experiments have revealed possible surface topological superconductivity with a critical temperature above 10 K and large spatial inhomogeneity in gap size (2-20 meV). In contrast, the surface superconducting gap reported in this work, comparable to the value in ARPES measurements (~1 meV)[32], is significantly smaller than those in previous STM works, despite the samples being prepared by the same self-flux method. The single crystals examined here display atomically flat cleavage surfaces extending over hundreds of nanometers, low defect density, robust bulk superconductivity, and well-defined surface Fermi arcs, unambiguously confirming their high quality. In the high-quality sample, our measurements provide compelling evidence for a second superconducting phase with an enhanced critical temperature and upper critical field, which remains homogeneous and robust within its domains. Notably, point-contact experiments[28,29] have also reported enhanced superconductivity, with a critical temperature of approximately 3 K, which has been attributed to possible electron doping or local strain. Unraveling the mechanism behind this phase requires further experimental and theoretical efforts.

Finally, the absence of magnetic vortices within surface superconducting regions, a feature not observed in earlier studies[35] or in the present work, has drawn considerable interest. Despite extensive measurements over multiple areas under various magnetic fields, no direct evidence of vortex formation was obtained. This may indicate a tip-related effect underlying the unusual behavior. Future studies employing alternative techniques, such as scanning quantum microscopy[48], are expected to provide further insight into this phenomenon.

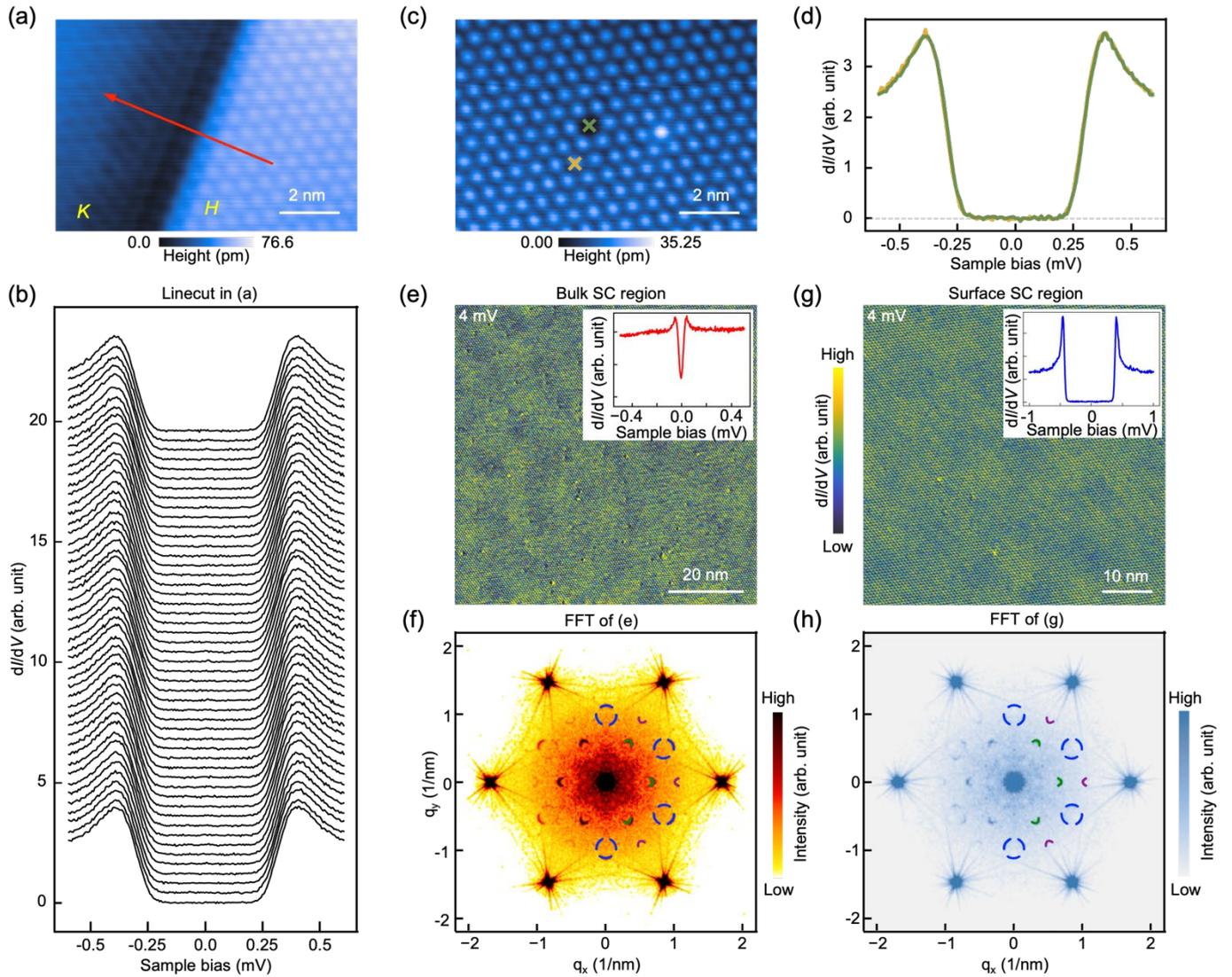

Fig. 4. The d$I$/d$V$ spectra across termination boundaries/defects and quasiparticle interference patterns at various surfaces of PtBi$_2$. (a) Topography of an edge between two types of termination surfaces. $V_s = -10$ mV, $I_t = 100$ pA. (b) Evolution of d$I$/d$V$ spectra along the red arrow in panel (a) across the edge, showing a uniform superconducting gap. (c) Topography of surface defect in H-type termination surface. (d) d$I$/d$V$ spectra taken at the positions marked by crosses with the same color in panel (c) under a magnetic field of 0.5 T. The d$I$/d$V$ spectrum taken on the defect (green) is almost identical to that away from the defect (orange). (e) d$I$/d$V$ map taken at 4 mV bias voltage on a flat H-type termination surface displaying the bulk superconductivity (inset). (f) Fourier transform of the d$I$/d$V$ map in the upper panel (e), showing signatures of surface Fermi arcs which are marked by blue, green and purple curves. (g) d$I$/d$V$ map taken at 4 mV in another area possessing the surface superconductivity (inset). (h) Fourier transform of the d$I$/d$V$ map from panel (g). Almost identical signatures of surface Fermi arcs, marked also by blue, green and purple curves, can be clearly resolved. Setpoint in (b-g): $V_s = -1$ mV, $I_t = 1$ nA, $V_{mod} = 0.005$ mV.

**Summary**

In summary, we present atomic-scale visualization of both bulk and surface superconductivity in the type-I Weyl semimetal γ-PtBi$_2$ using ultra-low-temperature (5 mK) scanning tunneling microscopy/spectroscopy. The bulk superconducting gap exhibits a size of Δ~0.053 meV, a critical temperature $T_c$~0.5 K, and an exceptionally low critical field <0.01 T. Zero-energy conductance mapping reveals well-defined vortices and vortex bound states, yielding a coherence length of approximately 200 nm. Remarkably, in certain surface regions a much larger gap (Δ~0.42 meV) emerges with a higher critical temperature ($T_c$~3 K) and a stronger critical field of ~2 T. Unlike the bulk phase, this surface gap shows no static vortex lattice and remains robust across different surface terminations and intrinsic defects. Quasiparticle interference measurements detect Fermi-arc scattering in both surface-superconducting and non-superconducting regions, indicating that Fermi arcs alone cannot account for the emergence of the surface superconductivity. These results establish γ-PtBi$_2$ as a rare platform hosting coexisting bulk and surface superconductivity and provide new insight into the mechanisms of unconventional superconductivity in topological materials.

During the preparation of the paper, we became aware that an independent similar work (arXiv:2508.04867) on the observation of a surface superconducting gap has been carried out by Moreno *et al.*[49].

*Experimental Method*

*Single-Crystal Growth of PtBi$_2$.* The single crystals of PtBi$_2$ were synthesized via the self-flux method[24]. The raw materials with molar ratio Pt:Bi = 1:8 were put into an alumina crucible, and then the crucible was sealed into a quartz tube under vacuum. The tubes were heated up to 800 °C, dwelled for 24 h, and then slowly cooled to 430 °C at a rate of 2 °C/h. At this temperature, the flux was separated by a centrifuge.

*Scanning tunneling microscopy/spectroscopy.* Experiments were performed in an ultrahigh vacuum ($1\times10^{-10}$ mbar) ultra-low temperature STM system equipped with an external magnetic field perpendicular to the sample surface. The ultra-low temperature is achieved by a dilution refrigerator with a cooling power of approximately 480 μW. The lowest base temperature measured at the mixing chamber stage is 5 mK. The electronic temperature is determined to be ~138 mK by calibrating on the Al(111) surface[50,51]. The PtBi$_2$ samples used in the STM/STS experiments were cleaved at room temperature (300 K) in an ultrahigh vacuum chamber. The samples were then transferred to the STM scanner and cooled down to 6 K. All the scanning parameters (setpoint voltage $V_s$ and tunneling current $I_t$) of the STM topographic images are listed in the figure captions. The d$I$/d$V$ spectra were acquired by a standard lock-in amplifier at a modulation frequency of 877.1 Hz. The modulation bias ($V_{mod}$) is listed in the figure captions. Non-magnetic tungsten tips were fabricated via electrochemical etching and calibrated on a clean Au(111) surface prepared by repeated cycles of sputtering with argon ions and annealing at 500 °C.

*Acknowledgement.* We thank Wenhan Dong, Xiaochun Huang, Hengxin Tan, Wei Li and Ziqiang Wang for helpful discussions. The work is supported by grants from the National Natural Science Foundation of China (62488201), the National Key Research and Development Projects of China (2022YFA1204100, 2023YFA1607400), the CAS Project for Young Scientists in Basic Research (YSBR-003) and the Innovation Program of Quantum Science and Technology (2021ZD0302700).

*Competing Interests.* The authors declare that they have no competing interests.